\documentclass[conference]{IEEEtran}
\IEEEoverridecommandlockouts
\usepackage{cite}
\usepackage{amsmath,amssymb,amsfonts}
\usepackage{algorithm}
\usepackage{algpseudocode}
\usepackage{graphicx}
\usepackage{textcomp}
\usepackage{xcolor}
\usepackage{mathtools}
\usepackage{tikz}
\usetikzlibrary{shapes,arrows}
\usepackage{subcaption}
\usepackage{xcolor,soul}
\usepackage{hyperref}
\usepackage{cases}
\input{mysymbol.sty}
\newtheorem{problem}{\hspace{-10pt}\bf Problem}

\renewcommand{\baselinestretch}{0.97} 

\def\BibTeX{{\rm B\kern-.05em{\sc i\kern-.025em b}\kern-.08em
    T\kern-.1667em\lower.7ex\hbox{E}\kern-.125emX}}
\begin{document}

\title{Detection by Sampling: Massive MIMO Detector based on Langevin Dynamics\\
\thanks{This work was partially supported by Nvidia. Email: \{nzilberstein, doost, ashu, segarra\}@rice.edu, cdick@nvidia.com.}
}

\author{%
  \IEEEauthorblockN{Nicolas Zilberstein$^{\star}$, Chris Dick$^{\dagger}$, Rahman Doost-Mohammady$^{\star}$, Ashutosh Sabharwal$^{\star}$, Santiago Segarra$^{\star}$}
  \IEEEauthorblockA{$^{\star}$Rice University, USA \hspace{4cm}
                    $^{\dagger}$Nvidia, USA} 
}

\maketitle
\begin{abstract}
Optimal symbol detection in multiple-input multiple-output (MIMO) systems is known to be an NP-hard problem. 
Hence, the objective of any detector of practical relevance is to get reasonably close to the optimal solution while keeping the computational complexity in check.
In this work, we propose a MIMO detector based on an {annealed} version of Langevin (stochastic) dynamics.
More precisely, we define a stochastic dynamical process whose stationary distribution coincides with the posterior distribution of the symbols given our observations.
In essence, this allows us to approximate the maximum a posteriori estimator of the transmitted symbols by sampling from the proposed Langevin dynamic.
Furthermore, we carefully craft this stochastic dynamic by gradually adding a sequence of noise with decreasing variance to the trajectories, which ensures that the estimated symbols belong to a pre-specified discrete constellation.
Through numerical experiments, we show that our proposed detector yields state-of-the-art symbol error rate performance.
\end{abstract}
%

\section{Introduction}\label{S:intro}
Massive multiple-input multiple-output (MIMO) systems are crucial for modern and future communications~\cite{mimoreview1}, \cite{mimoreview2}.
They are expected to play a key role in moving from the fifth to the sixth generation of cellular communications by achieving high data rates and spectral efficiency~\cite{6g}.
In massive MIMO systems, base stations are equipped with a large number of antennas, enabling them to handle several users simultaneously.
However, these systems entail many challenges such as designing low complexity MIMO detection schemes, which is the focus of our paper.

Exact MIMO detection is an NP-hard problem~\cite{Pia2017MixedintegerQP}. 
Given $N_u$ users and a modulation of $K$ symbols, the exact maximum likelihood (ML) estimator has an exponential decoding complexity $\mathcal{O}(K^{N_u})$.
Thus, obtaining this ML estimate is computationally infeasible and becomes intractable even for moderately-sized systems. 
Many approximate solutions for symbol detection have been proposed in the classical literature including zero forcing (ZF) and minimum mean squared error (MMSE)~\cite{Proakis2007}. 
Although both (linear) detectors have low complexity and good performance for small systems, their performance degrades severely for larger systems~\cite{chockalingam_rajan_2014}.
Another classical detector is approximate message passing (AMP), which is asymptotically optimal for large MIMO systems with Gaussian channels but degrades significantly for other (more practical) channel distributions~\cite{amp}. 
In the past few years, several massive MIMO symbol detectors based on machine learning -- and, in particular, deep learning -- have been derived.
One can roughly categorize them into channel-specific methods, like MMNet~\cite{mmnet}, and channel-agnostic methods like RE-MIMO~\cite{remimo}, OAMPNet~\cite{oampnet}, and hyperMIMO~\cite{zilberstein2021robust, hypermimo}.

An alternative family of detectors is based on Markov chain Monte Carlo (MCMC) methods~\cite[Chapter~8]{chockalingam_rajan_2014}.
Given that the ML estimator is prohibitive for large systems, these methods seek for a solution by generating candidate samples from the search space.
In particular, in \cite{MCMC_gibbs} a detector based on the Gibbs sampler was presented.
Recently, in~\cite{MIMOmetropolis}, the authors proposed a detector based on Metropolis-Hasting, in which they propose to make a random walk along the gradient descent direction of the least-square surface defined by the continuous-relaxed version of the ML.
In the past few years, in the context of image processing, sampling algorithms based on the Langevin dynamic have been proposed as generative models or to solve inverse problems. 
This iterative technique enables sampling from a given distribution by leveraging the availability of the score function (the gradient of the log-probability density function) without the necessity of computing the classical acceptance/rejection step in MCMC methods.
In~\cite{ermon2019}, an \textit{annealed} Langevin dynamic is used in the context of generative modeling for images.
Assuming an unknown distribution of the images, they parameterize the score function as a neural network and use the {annealed} Langevin dynamic to sample from the underlying probability distribution.
In \cite{kawar2021snips}, the authors proposed to solve noisy image inverse problems by sampling from the posterior.

In this work, we propose the first method that uses {annealed} Langevin dynamics for MIMO detection.
Given that the transmitted symbols come from a discrete constellation, we leverage the annealed process to include information of the prior in the dynamic.
As we have access to the prior distribution, we can define a closed-form expression for the score of the prior through the MMSE estimator.
This allows us to avoid the training process required in state-of-the-art learning-based detectors.
Hence, our detector can be applied to any observed channel and can handle a different number of users and mixed modulation schemes.

\vspace{0.7mm}
\noindent
{\bf Contribution.}
The contributions of this paper are twofold:\\
1) We propose a novel detector based on \emph{annealed} Langevin dynamics, allowing us to include information of our discrete prior in the exploration of the posterior distribution.\\
2) Through numerical experiments, we analyze the behavior of our method for different hyperparameter settings and demonstrate that the proposed detector achieves lower symbol error rate (SER) than baseline methods for massive MIMO systems.

\section{System model and problem formulation}

We consider a communication channel with $N_u$ single-antenna transmitters or users and a receiving base station with $N_r$ antennas. 
The forward model for this MIMO system is given by
\begin{equation}\label{E:mimo_model}
	\bby = \bbH \bbx + \bbz,
\end{equation}
where $\bbH \in \mathbb{C}^{N_r \times N_u}$ is the channel matrix, $\bbz \sim \mathcal{CN}(\bb0, \sigma_0^2 \bbI_{N_r})$ is a vector of complex circular Gaussian noise, $\bbx \in \mathcal{X}^{N_u}$ is the vector of transmitted symbols, $\mathcal{X}$ is a finite set of constellation points, and $\bby \in \mathbb{C}^{N_r}$ is the received vector. 
In this work, a quadrature amplitude modulation (QAM) is used and symbols are normalized to attain unit average power. 
It is assumed that the constellation is the same for all transmitters and each symbol has the same probability of being chosen by the users $N_{u}$. 
Moreover, perfect channel state information (CSI) is assumed, which means that $\bbH$ and $\sigma_0^2$ are known at the receiver.\footnote{To avoid notation overload, we adopt the convention that whenever we assume $\bbH$ to be known, $\sigma_0^2$ is also known.} 
Under this setting, the MIMO detection problem can be defined as follows.

\vspace{2mm}
\begin{problem}\label{P:main} \emph{
	Given perfect CSI and an observed $\bby$ following~\eqref{E:mimo_model}, find an estimate of $\bbx$.}
\end{problem}
\vspace{2mm}

Given the stochastic nature of $\bbz$ in~\eqref{E:mimo_model}, a natural way of solving Problem~\ref{P:main} is to search for the $\bbx$ that maximizes its \textit{posterior} probability given the observations $\bby$.
Hence, the optimal decision rule can be written as
\begin{align}\label{eq:map}
	\hat{\bbx}_{\mathrm{MAP}} &= \argmax_{\bbx \in \mathcal{X}^{N_u}}\,\, p(\bbx|\bby,\bbH)\\
	&= \argmax_{\bbx \in \mathcal{X}^{N_u}}\,\, p_{\bbz}(\bby - \bbH\bbx)p(\bbx)\nonumber,
\end{align}
%
\noindent where we have applied Bayes' rule. 
As we assume that the symbols' prior distribution is uniform among the constellation elements and the measurement noise $\bbz$ is Gaussian, the maximum a posteriori (MAP) detector reduces to an ML detector. 
Specifically, \eqref{eq:map} boils down to solving the following optimization problem
\begin{equation}\label{eq:ml}
	\hat{\bbx}_{\mathrm{ML}} = \argmin_{\bbx \in \mathcal{X}^{N_u}}\,\, ||\bby - \bbH\bbx||^2_2,
\end{equation}
%

\noindent which is NP-hard due to the finite constellation constraint $\bbx \in \mathcal{X}^{N_u}$, rendering $\hat{\bbx}_{\mathrm{ML}}$ intractable in practical applications.
Consequently, several schemes have been proposed in the last decades to provide efficient approximate solutions to Problem~\ref{P:main}, as mentioned in Section~\ref{S:intro}.
In this paper, we propose to solve Problem~\ref{P:main} by (approximately) sampling from the posterior distribution in~\eqref{eq:map} using an annealed Langevin dynamic.

\section{Langevin for MIMO detection}

In Section~\ref{subsec:langevindyn} we briefly introduce the Langevin dynamic while in Section~\ref{subsec:posterior} we explain how we propose to use it for MIMO detection.
In particular, we detail the expressions of the score functions involved in the sampling process to solve our Problem~\ref{P:main}.

\subsection{Langevin dynamics}
\label{subsec:langevindyn}

The Langevin dynamic algorithm is an MCMC algorithm \cite{MCMCbook, Roberts1996ExponentialCO}, described by the following equation\footnote{This is technically known as unadjusted Langevin algorithm (ULA), which is obtained from the Euler-Maruyama discretization of the overdamped Langevin stochastic differential equation~\cite{durmus_moulines}.}

\begin{equation}\label{eq:langevin}
	\bbx_{t+1} = \bbx_t + \epsilon \nabla_{\bbx_t}\log p(\bbx_t) + \sqrt{2\epsilon}\, \bbw_t,
\end{equation}
where $p(\bbx)$ is the target distribution from which we want to generate samples $\bbx \in \mathbb{R}^N$ and $\bbw_t \sim \ccalN(0, \bbI_N)$. 
The dynamic in~\eqref{eq:langevin} explores the target distribution by moving in the direction of the gradient of the logarithm of the target density $\nabla_{\bbx}\log p(\bbx)$, known as \textit{score function}.
In essence, it is a combination of stochastic gradient ascent in the direction of the score function and injected noise, which allows the method to avoid collapsing to local maxima.
Under some regularity conditions~\cite{wellinglang}, the distribution of $\bbx_T$ is equal to $p(\bbx)$ when $\epsilon \rightarrow 0$ and $T \rightarrow \infty$, in which case $\bbx_T$ becomes an exact sample from $p(\bbx)$. 
In practice, neither $\epsilon \rightarrow 0$ nor $T \rightarrow \infty$, so a Metropolis-Hastings acceptance/rejection step is used to ensure convergence, leading to the so-called Metropolis-adjusted Langevin algorithm (MALA)~\cite{Roberts1996ExponentialCO}.
An alternative, proposed in~\cite{wellinglang}, is to use a time-inhomogeneous variant of~\eqref{eq:langevin}, i.e., defining a variable step size $\epsilon_t$.
They demonstrate that when $\epsilon_t$ decreases to zero for large $t$, the error becomes negligible and the acceptance/rejection step can be omitted.
It should be noted that the \textit{only} requirement for sampling from $p(\bbx)$ using this procedure is knowing the score function.

\subsection{Detection by sampling from the posterior distribution}
\label{subsec:posterior}

Given the intractability of Problem~\ref{P:main} due to the finite constellation constraint, we propose to \emph{generate a set of samples that approximately come from the posterior distribution $p(\bbx|\bby, \bbH)$ using~\eqref{eq:langevin} and then select the one that minimizes the objective in~\eqref{eq:ml}}.
The key ingredient in the Langevin dynamic is the score function, which for our case is given by $\nabla_{\bbx}\log p(\bbx| \bby, \bbH)$.
After applying Bayes' rule, this score function can be rewritten as 
\begin{equation}\label{E:score_function}
\nabla_{\bbx}\log p(\bbx|\bby, \bbH) = \nabla_{\bbx}\log p(\bby|\bbx,\bbH) + \nabla_{\bbx}\log p(\bbx),
\end{equation}
where the term $\nabla_{\bbx}\log p(\bby|\bbx,\bbH)$ corresponds to the score function of the likelihood and $\nabla_{\bbx}\log p(\bbx)$ to the score function of the prior.
Notice that this latter term is not well defined due to the discrete nature of the symbols.

To circumvent this obstacle, inspired by~\cite{kawar2021snips}, we approximate the prior by using an \textit{annealed} version of the Langevin dynamic. 
Specifically, instead of working with the discrete symbols $\bbx$, we define a perturbed version of the symbols $\tilde{\bbx} = \bbx + \bbn$ with $\bbn \sim \mathcal{CN}(0, \sigma^2\bbI)$, for different values of $\sigma^2$. 
Note that $\tilde{\bbx}$ now has a continuous prior allowing us to run a Langevin dynamic in $\tilde{\bbx}$ instead of $\bbx$.
Moreover, if we make $\sigma^2 \to 0$ then $\tilde{\bbx}$ concentrates around $\bbx$, allowing us to effectively sample from $p(\bbx|\bby, \bbH)$, as wanted.

In a nutshell, the algorithm works as follows. 
First, we initialize $\tilde{\bbx}_0$ uniformly at random in $[-1,1] \times [-1,1]$, since the symbols are assumed to be normalized. 
Then, we follow the direction of the score function of the log-posterior density of the perturbed symbol $\nabla_{\tilde{\bbx}}\log p(\tilde{\bbx}| \bby, \bbH)$, starting with a high $\sigma$ and gradually decreasing its value until $\tilde{\bbx} \approx \bbx$.
Apart from enabling the approximation of the score function of the prior distribution, the annealing process also improves the mixing time of the Langevin dynamic~\cite{ermon2019}.
This is particularly important in multimodal distributions, as is the case of MIMO detection.
Having introduced the high-level idea of our method, we now provide more details on the \emph{annealing process}, exact expressions for the terms in the \emph{score function}, and a step-by-step description of the \emph{algorithm}.

\vspace{1mm}

\begin{figure}[t]
	\centering
	\includegraphics[width=0.4\textwidth]{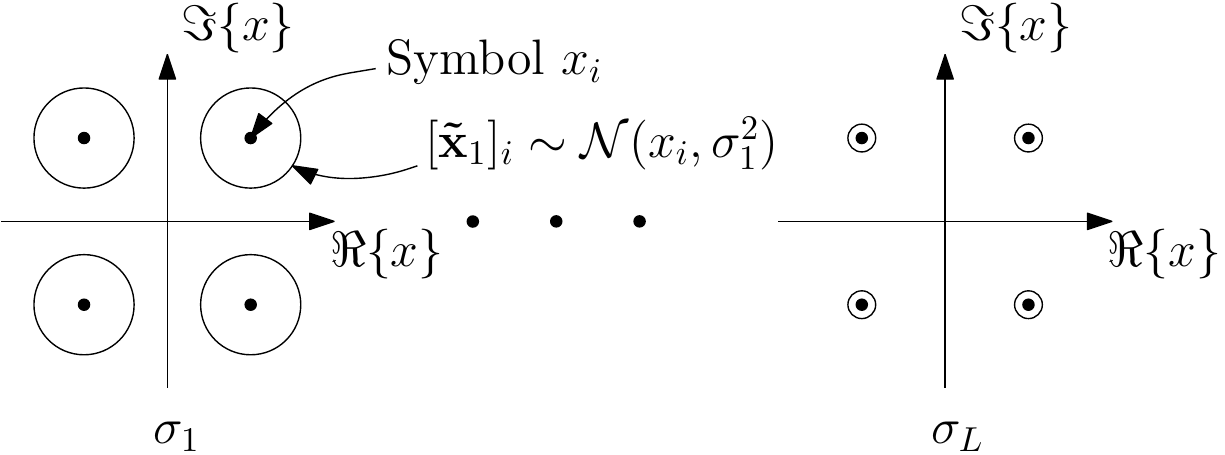}
	\vspace{-0.03in}
	\caption{{\small Scheme of the annealed process. We consider a QPSK constellation and at each level we add Gaussian noise. The variance of the noise decreases for higher levels. In the last level $L$, the Gaussian is very sharp around each symbol, mimicking our true discrete prior over the constellation.}}
	\vspace{-0.1in}
	\label{fig:gaussian_levels}
\end{figure}

\noindent{\bf Annealing process.}
We define a sequence of noise levels $\{\sigma_l\}_{l=1}^{L}$ such that $\sigma_1 > \sigma_2 > \cdots > \sigma_L > 0$. 
Then, at each level we define a perturbed version of the true symbols $\bbx$
\begin{equation}\label{eq:pert_symbs}
    \tilde{\bbx}_{l} = \bbx + \bbn_{l},
\end{equation}
where $\bbn_{l} \sim \mathcal{CN}(0, \sigma_l^2\bbI)$. 
A representation (for a QPSK modulation) of this process is shown in Fig. \ref{fig:gaussian_levels}.
Since the variance of the noise injected at each level is a predefined parameter, we will design the sequence in such way that the noise injected in the last levels is very small, approximating the true discrete distribution given by a set of delta functions at each symbol with uniform weight.

\vspace{1mm}
\noindent {\bf Score function.}
Given the perturbed symbols in~\eqref{eq:pert_symbs}, the forward model in~\eqref{E:mimo_model} can be rewritten as
\begin{align}
	\bby &= \bbH\tilde{\bbx}_l + (\bbz - \bbH\bbn_l).
\end{align}
In this new forward model, the likelihood is given by $p(\bby|\tilde{\bbx}_l, \bbH) = p(\bbz - \bbH {\bbn}_l|\tilde{\bbx}_l)$, which is not Gaussian: although $\bbn_l$ is a Gaussian random variable, when conditioning on $\tilde{\bbx}_l$ the conditional distribution is no longer Gaussian due to~\eqref{eq:pert_symbs}.
However, an analytical expression for the score of the likelihood can still be obtained by following the approach in~\cite{kawar2021snips}, where a synthetic annealed noise carved from the measurement noise $\bbz$ is constructed in a gradual fashion.
Furthermore, in order to get a tractable expression, we have to rely on the singular value decomposition (SVD) of the channel matrix given by $\bbH = \bbU\boldsymbol{\Sigma}\bbV^\top$ as well as in the spectral representation of $\tilde{\bbx}_l$ and $\bby$ defined as $\tilde{\boldsymbol{\chi}}_l = \bbV^{\top} \tilde{\bbx}_l$ and $\boldsymbol{\eta} = \bbU^{\top}\bby$.
In essence, the gradual noise addition is constructed in such a way that the noise $\bbz - \bbH\bbn_l$ is uncorrelated and independent of $\tilde{\bbx}_l$, given the singular values $s_j = [\boldsymbol{\Sigma}]_{jj}$. 
More precisely, the distribution of $\bbz - \bbH {\bbn}_l$ is given by a multivariate Gaussian distribution, where each component is distributed as $[\bbz - \bbH {\bbn}_l]_j \sim \ccalN(0, |\sigma_0^2 - \sigma_l^2s_j^2|)$ for $j=1,\cdots, N_u$.

With this spectral representation in mind, our goal is to run a Langevin dynamic whose score function for every noise level $l$ is given by [cf.~\eqref{E:score_function}]
\begin{equation}\label{E:score_function_spectral}
\nabla_{\tilde{\boldsymbol{\chi}}_l}\!\log p(\tilde{\boldsymbol{\chi}}_l| \boldsymbol{\eta}, \bbH) = \nabla_{\tilde{\boldsymbol{\chi}}_l} \log p(\boldsymbol{\eta}|\tilde{\boldsymbol{\chi}}_l, \bbH) + \nabla_{\tilde{\boldsymbol{\chi}}_l}\log p(\tilde{\boldsymbol{\chi}}_l).
\end{equation}
We now provide closed-form expression for both constituent terms in this score function.
\vspace{1mm}

\noindent \emph{i) Score of the likelihood:} Given the above discussion, the final expression for the score of the likelihood in the spectral domain is given by
\begin{equation}\label{eq:score_likeli}
    \nabla_{\tilde{\boldsymbol{\chi}}_l} \! \log p(\boldsymbol{\eta}|\tilde{\boldsymbol{\chi}}_l, \bbH)  = \boldsymbol{\Sigma}^\top \,\, |\sigma_0^2\bbI - \sigma_l^2\boldsymbol{\Sigma}\boldsymbol{\Sigma}^\top|^{\dagger}\,\, ( \boldsymbol{\eta} - \boldsymbol{\Sigma} \tilde{\boldsymbol{\chi}}_l).
\end{equation}
To give some intuition, the score function of the likelihood is given by the gradient of a multivariate Gaussian distribution: the residual error $( \boldsymbol{\eta} - \boldsymbol{\Sigma} \tilde{\boldsymbol{\chi}}_l) = (\bbU^\top\bby - \boldsymbol{\Sigma}\bbV^\top \tilde{\bbx}_l)$ is multiplied by the (pseudo-)inverse of the covariance matrix, which is diagonal with entries given by $|\sigma_0^2 - \sigma_l^2s_j^2|$. 
For details about the derivation of this expression see~\cite{kawar2021snips}.

\vspace{1mm}
\noindent \emph{ii) Score of the annealed prior:} We first notice that $\nabla_{\tilde{\boldsymbol{\chi}}_l}\log p(\tilde{\boldsymbol{\chi}}_l) = \bbV^{\top}\nabla_{\tilde{\bbx}_l}\log p(\tilde{\bbx}_l)$ due to the orthogonality of $\bbV$.
Moreover, based on the Tweedie's identity~\cite{TweedieIdent}, we can relate the score function $\nabla_{\tilde{\bbx}_l}\log p(\tilde{\bbx}_l)$ and the MMSE denoiser as follows
\begin{equation}\label{eq:prior}
	\nabla_{\tilde{\bbx}_l}\log p(\tilde{\bbx}_l) = \frac{\mathbb{E}_{\sigma_l}[\bbx|\tilde{\bbx}_l] - \tilde{\bbx}_l}{\sigma_l^2}.
\end{equation}
In particular, the conditional expectation can be calculated elementwise as
\begin{align}\label{E:mixed_gaussian}
	\mathbb{E}_{\sigma_l}[x_j|[\tilde{\bbx}_l]_j] &= \frac{1}{Z}\sum_{x_k \in \ccalX} x_k \exp\bigg(\frac{-||[\tilde{\bbx}_l]_j - x_k||^2}{2\sigma_l^2}\bigg),
\end{align}
where $Z = \sum_{x_k \in \ccalX} \exp\Big(\frac{-||[\tilde{\bbx}_l]_j - x_k||^2}{2\sigma_l^2}\Big)$ and $j=1,\cdots, N_u$. 

\vspace{2mm}
\noindent {\bf Algorithm.} 
The algorithm to generate samples $\hat{\bbx}$ from the (approximate) posterior $p(\bbx|\bby,\bbH)$ is shown in Alg.~\ref{alg}.
As discussed in~\cite{kawar2021snips}, when computing the entries of the score function in~\eqref{E:score_function_spectral} using the expressions in~\eqref{eq:score_likeli} and~\eqref{eq:prior}, one of these terms might be negligible with respect to the other depending on the noise level.
Thus, the elementwise score of the posterior will be given by
 \begin{align}\label{eq:full_score}
 [\nabla_{\tilde{\boldsymbol{\chi}}_l} & \! \log (\boldsymbol{\eta}|\tilde{\boldsymbol{\chi}}_l, \bbH)]_j = \\
 & 
    \begin{cases}
  [\nabla_{\tilde{\boldsymbol{\chi}}_l} \! \log p(\boldsymbol{\eta}|\tilde{\boldsymbol{\chi}}_l, \bbH) + \bbV^{\top}\nabla_{\tilde{\bbx}}\log p(\tilde{\bbx})]_j, \,\,\,\, \sigma_0 \geq \sigma_ls_j
  \\
  [\nabla_{\tilde{\boldsymbol{\chi}}_l} \! \log p(\boldsymbol{\eta}|\tilde{\boldsymbol{\chi}}_l, \bbH)]_j,  \hspace{29mm} \sigma_0 < \sigma_ls_j \\
  [\bbV^{\top}\nabla_{\tilde{\bbx}} \! \log p(\tilde{\bbx})]_j, \hspace{34mm}  s_j = 0.
\end{cases}
\nonumber
\end{align}
Intuitively, when the injected noise at level $l$ is such that $\sigma_l s_j > \sigma_0$, then the contribution of the score of the prior is negligible and can be ignored.
Similarly, whenever $s_j=0$, the corresponding entry $\eta_j$ is uninformative and the score of the likelihood can be ignored.
Furthermore, a refinement that we incorporate in the algorithm is the use of position-dependent step sizes.
Instead of using a constant \emph{scalar} step size $\epsilon$ as in~\eqref{eq:langevin} or even time-varying versions of it, in Alg.~\ref{alg} we employ level-dependent \emph{diagonal matrices} $\boldsymbol{\Lambda}_l$.
In this way, different entries of our vector-valued Langevin dynamic can be updated at different rates depending on the singular values of the channel under consideration.
Finally, given that $\sigma_L \neq 0$, the sample will be very close to the constellation but not exactly. Hence, we take $\hat{\bbx} = \argmin_{\bbx \in \ccalX^{N_u}}||\bbx - \bbV\tilde{\boldsymbol{\chi}}_{T,L}||_2^2$.

Given that Alg.~\ref{alg} is stochastic, one can generate several samples $\hat{\bbx}$ from the same (approximate) posterior distribution by running the algorithm multiple times. 
Therefore, as we want to approximate the MAP estimate -- equivalently for this case, the ML estimate -- we run $M$ different Langevin trajectories for each pair $\{\bby, \bbH\}$ and keep the sample that minimizes~\eqref{eq:ml}.
Formally, given $M$ samples $\{\hat{\bbx}_m\}_{m=1}^M$ obtained from Alg.~\ref{alg}, our final estimate is given by
\vspace{-0.2cm}
\begin{equation}\label{eq:Ntraj}
    \hat{\bbx} = \argmin_{\bbx \in \{\hat{\bbx}_m\}_{m=1}^M} ||\bby - \bbH\bbx||_2^2.
\end{equation}
\vspace{-0.3cm}

\noindent Notice that these $M$ trajectories can be run in parallel, as they are independent of each other.

\vspace{1mm}
\noindent {\bf Computational complexity.} 
The first step in Alg.~\ref{alg} is to compute the SVD of the channel $\bbH$, whose complexity is $\ccalO(N_uN_r\min\{N_u, N_r\})$, and is done only once per channel.
Moreover, notice that the matrices involved in each iteration are diagonal so the complexity of multiplying them is $\ccalO(N_u^2)$, while the matrix inversion is $\ccalO(N_u)$.
Finally, given a modulation of $K$ symbols, the complexity of~\eqref{E:mixed_gaussian} is $\ccalO(KN_u)$.
Hence, one iteration has a complexity of $\ccalO(N_u^2 + KN_u)$.
The overall complexity, including the SVD computation and all the iterations, is $\ccalO(N_uN_r\min\{N_u, N_r\} + LT(N_u^2 + KN_u))$.
Regarding the $M$ trajectories, observe that these are independent of each other, so they can be computed in parallel.
Therefore, the bottlenecks are twofold: the SVD computation and the number of iterations $LT$. 
While the former is inevitable, the latter is a parameter that we control and represents a trade-off between SER performance and computational complexity.
In Section~\ref{sec:results}, we present some numerical experiments that analyze this trade-off and the impact on the SER performance.

\begin{algorithm}[t]
	\caption{Annealed Langevin for MIMO detection}\label{alg}
	\begin{algorithmic}
		\Require $T, \{\sigma_l\}_{l=1}^L, \epsilon, \sigma_0, \bbH, \bby$
		\State Compute SVD of $\bbH = \bbU\boldsymbol{\Sigma}\bbV^{\top}$
		\State Initialize $\tilde{\boldsymbol{\chi}}_{t=0,l=1}$ with random noise $\ccalU[-1,1]$
		\For{$l = 1\; \text{to}\;  L$}
		\State $[\boldsymbol{\Lambda}_l]_{jj} =$
		$\begin{cases}
		  \frac{\epsilon \sigma_l^2 }{\sigma_L} (1 - \frac{\sigma_l^2}{\sigma_0^2}s_j^2) \hspace{8mm} \text{if} \,\,\, \sigma_ls_j \leq \sigma_0 \\
		\frac{\epsilon}{\sigma_L} (\sigma_l^2 - \frac{\sigma_0^2}{s_j^2}) \hspace{10mm}  \text{if} \,\,\, \sigma_ls_j > \sigma_0
		\end{cases}$
		
		\For{$t = 0\; \text{to}\; T-1$}
			\State Draw $\bbw_t \sim \ccalN(0, \bbI)$
			
            \State Compute $\nabla_{\tilde{\boldsymbol{\chi}}_{t,l}}\log p(\boldsymbol{\eta}|\tilde{\boldsymbol{\chi}}_{t,l}, \bbH)$ as in~\eqref{eq:score_likeli}
            
            \State Compute $\nabla_{\tilde{\bbx}_{t,l}}\log p(\tilde{\bbx}_{t,l})$
            as in~\eqref{eq:prior}
            
            \State Compute $\nabla_{\tilde{\boldsymbol{\chi}}_{t,l}}\!\log p(\tilde{\boldsymbol{\chi}}_{t,l}| \boldsymbol{\eta}, \bbH)$ as in~\eqref{eq:full_score}

            
            \State $\tilde{\boldsymbol{\chi}}_{t+1, l} = \tilde{\boldsymbol{\chi}}_{t,l} + \boldsymbol{\Lambda}_l \nabla_{\tilde{\boldsymbol{\chi}}_{t,l}}\!\log p(\tilde{\boldsymbol{\chi}}_{t,l}| \boldsymbol{\eta}, \bbH) + \sqrt{2\boldsymbol{\Lambda}_l}\, \bbw_t$
		\EndFor
		\State $\tilde{\boldsymbol{\chi}}_{0, l+1} = \tilde{\boldsymbol{\chi}}_{T, l}$
		\EndFor \\
	\Return $\hat{\bbx} = \argmin_{\bbx \in \ccalX^{N_u}}||\bbx - \bbV\tilde{\boldsymbol{\chi}}_{T,L}||_2^2$
	\end{algorithmic}
\end{algorithm}

\section{Results}\label{sec:results}
In this section we present the results of our proposed method.\footnote{Code to replicate the numerical experiments can be found
at \url{https://github.com/nzilberstein/Langevin-MIMO-detector}}
We start by presenting the channel model and the simulation setup.
Then, we analyze the SER performance of the proposed method when considering different noise levels $L$ and different numbers of trajectories $M$.
Finally, we compare our method with both classical and learning-based baseline detectors.

\vspace{1mm}
\noindent {\bf Channel model and simulation settings.} The channel model is generated following the Kronecker correlated model
\begin{equation}
	\bbH = \bbR_r^{1/2}\bbH_e \bbR_u^{1/2},
	\label{E:kron}
	\vspace{-0.07in}
\end{equation}
where $\bbH_e$ is a Rayleigh fading channel matrix and $\bbR_r$ and $\bbR_u$ are the spatial correlation matrices at the receiver and transmitters, respectively, generated according to the exponential correlation matrix model with a correlation coefficient $\rho$; see \cite{Loyka2001} for details.
The signal-to-noise ratio (SNR) is given by 
\begin{equation}
	\text{SNR} = \frac{\mathbb{E}[||\bbH\bbx||^2]}{\mathbb{E}[||\bbz||^2]} = \frac{N_u}{\sigma_0^2N_r}.
	\vspace{-0.05in}
\end{equation}
The simulation environment includes a base station with $N_r=64$ receiver antennas and $N_u=32$ single-antenna users.
We consider a 16-QAM modulation and $\rho = 0.6$. 
The value of $\epsilon$ is fixed at $3 \times 10^{-5}$, while the number of samples per noise level at $T=70$.
The batch size for testing is $5000$.

\begin{figure*}[t]
	\begin{subfigure}{.33\textwidth}
    	\centering
    	\includegraphics[width=1\textwidth]{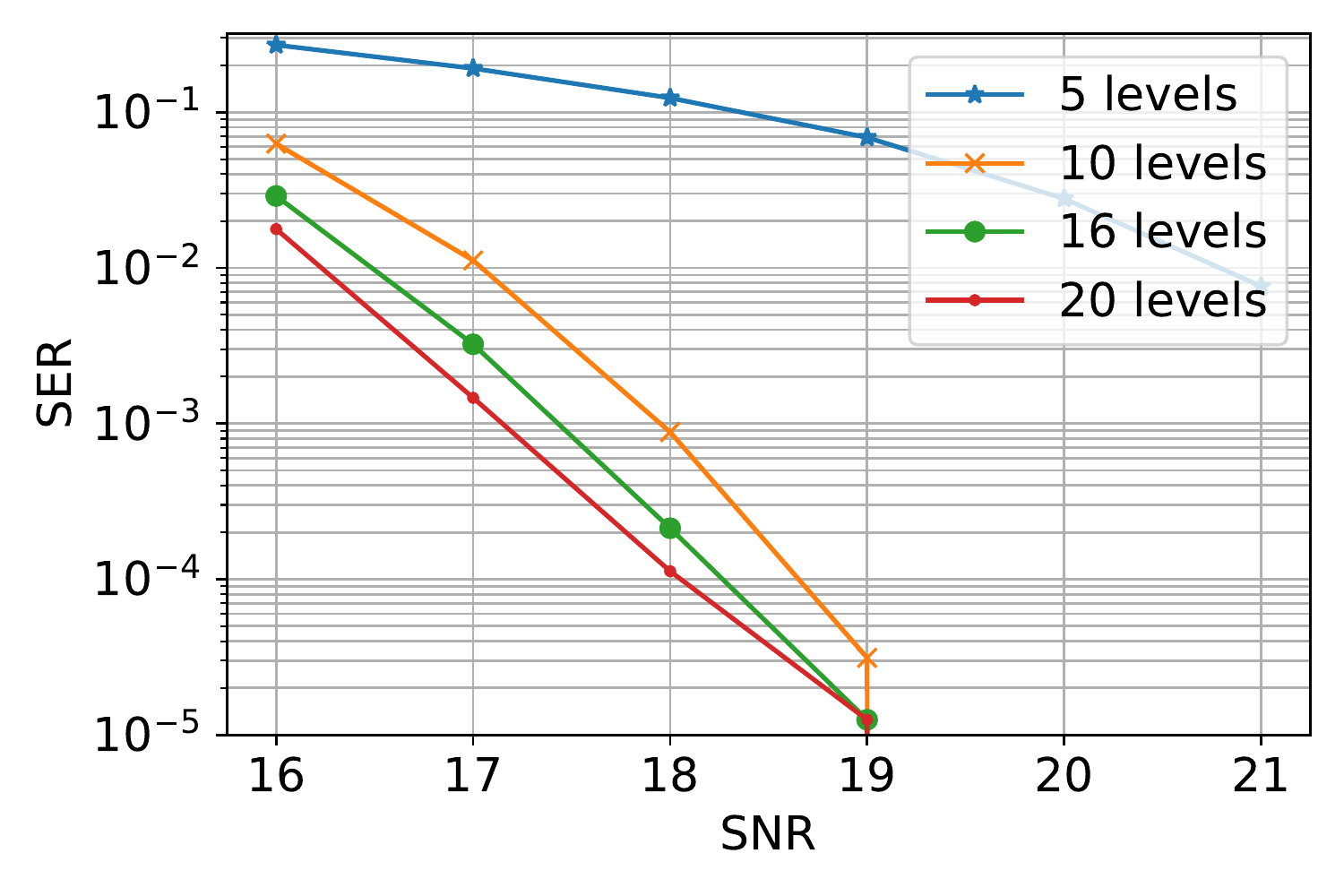}
    	\vspace{-0.15in}
    	\caption{}
    	\label{fig:SER-noiselevels}
	\end{subfigure}%
	\begin{subfigure}{.33\textwidth}
    	\centering
    	\includegraphics[width=1\textwidth]{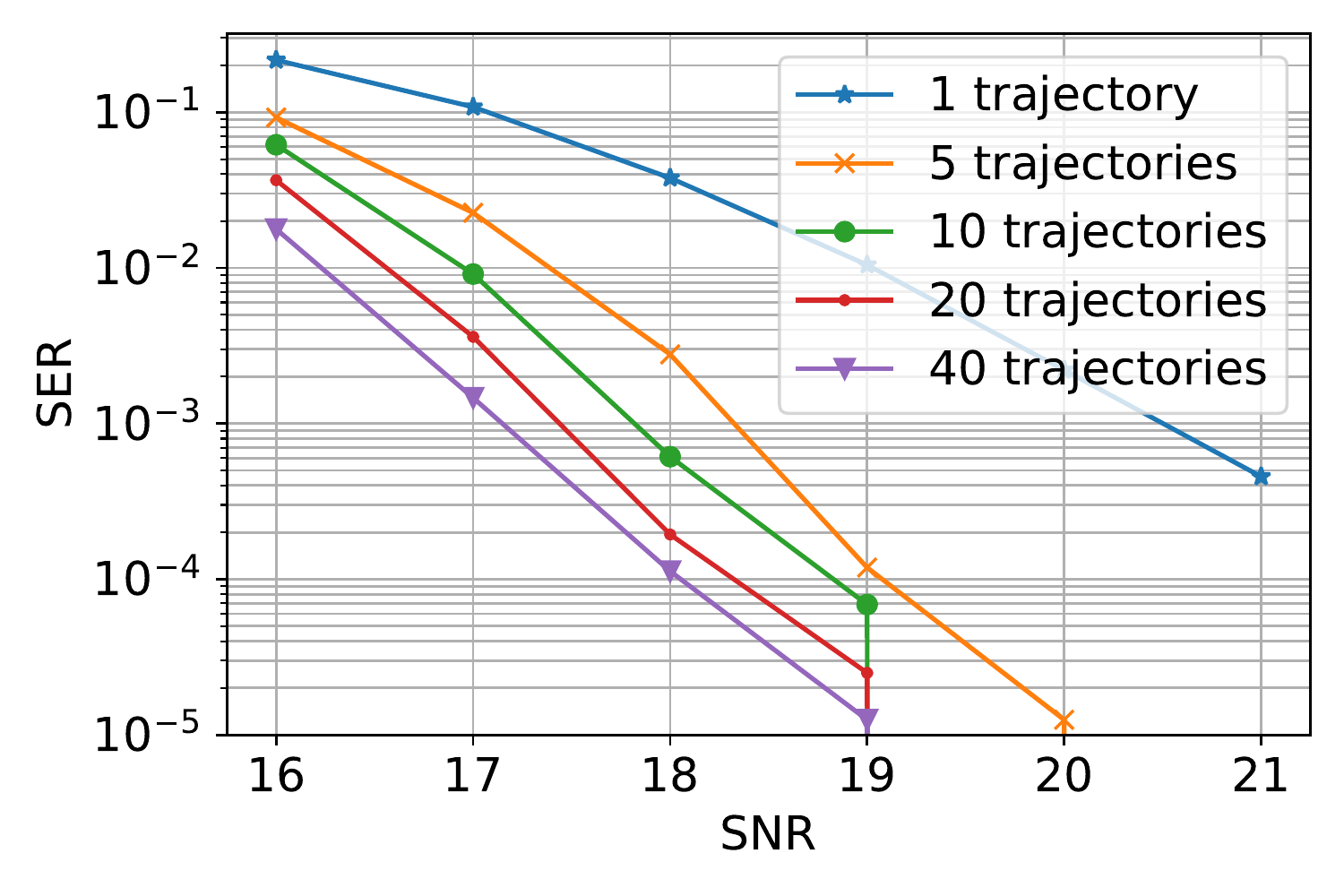}
    	\vspace{-0.15in}
    	\caption{}
    	\label{fig:SER-sertraj}
	\end{subfigure}
	\begin{subfigure}{.33\textwidth}
    	\centering
    	\includegraphics[width=1\textwidth]{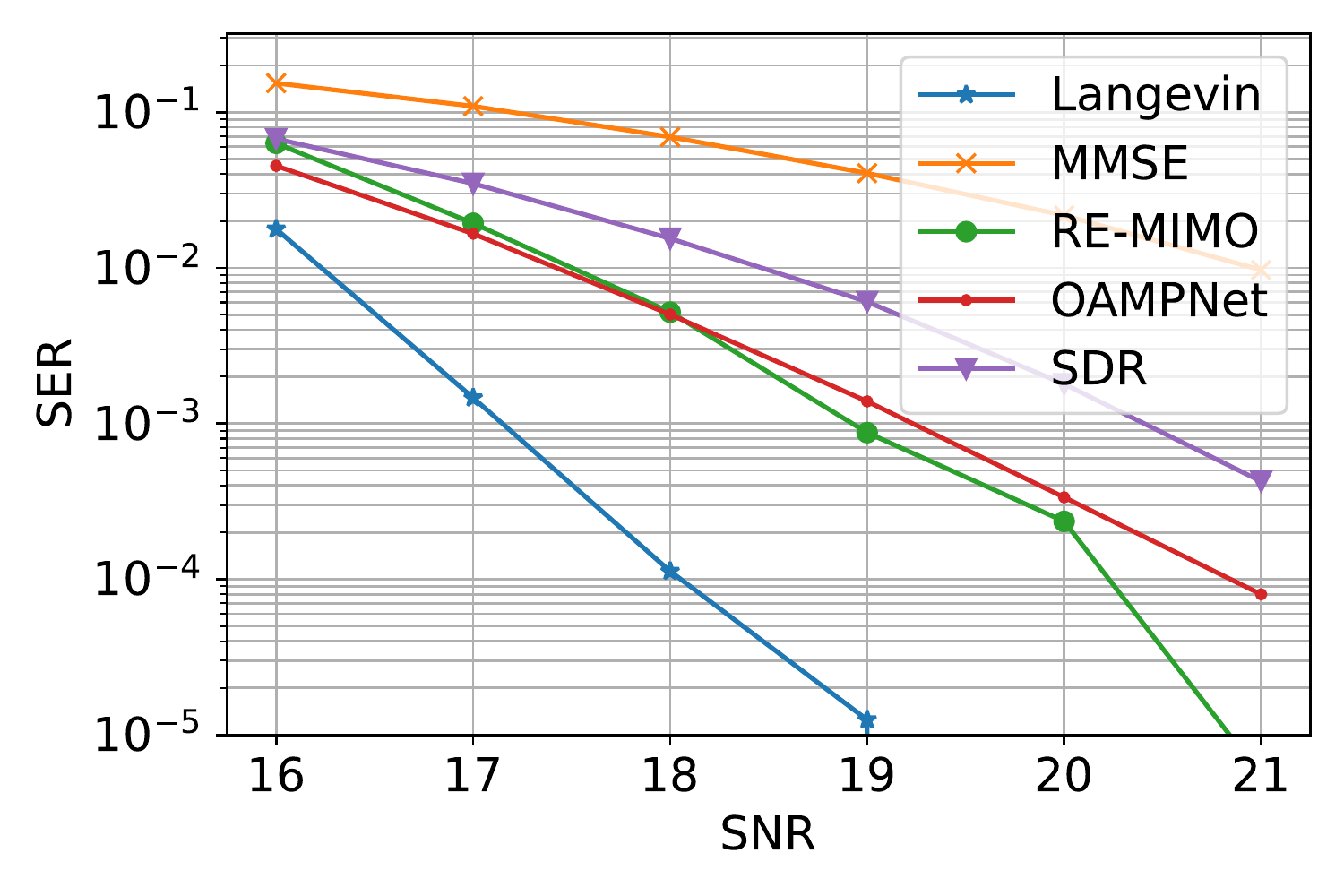}
    	\vspace{-0.15in}
    	\caption{}
    	\label{fig:SER-comparison}
	\end{subfigure}
	\vspace{-0.02in}
	\caption{ {\small Performance analysis of our proposed method. (a)~SER as a function of SNR for our Langevin method for noise levels $L \in \{5, 10, 16, 20\}$. (b)~SER as a function of SNR for our Langevin method for $M \in \{1, 5, 10, 20, 40\}$ numbers of trajectories. (c)~SER as a function of SNR for different detection methods evaluated in a Kronecker correlated channel model as in~\eqref{E:kron}.}}
	\vspace{-0.1in}
	\label{fig_results}
\end{figure*}

\vspace{1mm}
\noindent{\bf Varying the number of noise levels.} 
In the first experiment, given the sequence of noise levels with variance $\{\sigma_l\}_{l=1}^L$, we fix $\sigma_1 = 1$, $\sigma_L = 0.01$ and $M=40$ trajectories and change the number of noise levels between them.
We consider four cases where $L \in \{5, 10, 16, 20\}$; see Fig.~\ref{fig:SER-noiselevels}.
First, notice that the performance when $L=5$ is much worse than the other three cases.
This implies that the algorithm is not able to sufficiently explore the search space. 
On the other hand, the gap between the other three cases is much smaller.
Therefore, a trade-off between computational burden and performance exists: when considering more levels, the SER performance improves at the cost of increasing the running time.
From this experiment, we conclude that at least $L=10$ levels are needed in order to perform as well as the existing state-of-the-art detectors.

\vspace{1mm}
\noindent{\bf Varying the number of trajectories.} 
The performance of the detector as a function of the number $M$ of different Langevin trajectories [cf.~\eqref{eq:Ntraj}] is shown in Fig.~\ref{fig:SER-sertraj}.
We analyze five cases where $M \in \{ 1, 5, 10, 20, 40 \}$.
For all the cases, we consider $L=20$ noise levels.
From the results, we see that $M$ is a hyperparameter that has a high impact on the overall performance of the detector. 
In particular, if we consider only $M=1$, then the performance degrades severally, with a SER in the order of the classical MMSE detector (not shown in the figure).
However, if we consider $M=40$ trajectories, the proposed method outperforms state-of-the-art detectors, as we illustrate in our next experiment.

\vspace{1mm}
\noindent{\bf Performance comparison with baseline methods.}
Based on our previous experiments, we set $L=20$ noise levels between $\sigma_1 = 1$ and $\sigma_{20} = 0.01$, and we run $M=40$ trajectories.
We compare our method with the following detectors: MMSE detector, semidefinite relaxation detector (SDR)~\cite{SDR1}, and two  learning-based, RE-MIMO~\cite{remimo} and OAMPNet~\cite{oampnet}, which were trained as explained in the respective papers with channels drawn from~\eqref{E:kron}.
The comparison is shown in Fig.~\ref{fig:SER-comparison}.
The figure reveals that our proposed method markedly outperforms the other detectors.
It is particularly interesting to notice that our proposed method outperforms the learning-based detectors, which have been trained with channel instances drawn from the same model as those in the testing set.
Moreover, since our method does not require training, it is very flexible and presents two main advantages over the learning-based baselines: it can handle channels drawn from any distribution (promoting its application to real-world channel instances) and a varying number of users without the need of any retraining as required in, e.g., OAMPNet~\cite{oampnet}.
This is key in MIMO communications, as the number of users connected to the network is constantly changing.
%
\section{Conclusions}
\vspace{-0.6mm}
We proposed a massive MIMO detector based on an {annealed} version of Langevin dynamics that achieves state-of-the-art SER performance on correlated channels in large-scale systems. 
To include the prior information in the sampling process, we approximated the discrete prior distribution with a sequence of annealed noises that tend to concentrate around the discrete constellation symbols.
Future work includes running experiments for scenarios where users are transmitting with multiple modulation schemes simultaneously, extending our method to scenarios with imperfect CSI, and leveraging the rich Langevin theory to derive theoretical guarantees.
\vspace{-1mm}






\bibliographystyle{IEEEbib}
\bibliography{citations}

\end{document}